\documentclass[prl,twocolumn,showpacs,groupedaddress,citeautoscript]{revtex4}
\usepackage{graphicx}
\usepackage{amsmath,amssymb}
\graphicspath{{./images/}}

\newcommand{\tr}{\mathop{\mathrm{tr}}}
\renewcommand{\section}[1]{{\par\it #1.---}\ignorespaces}

\begin{document}
\title{Time-Resolved Measurement of a Charge Qubit}
 \author{Georg M. Reuther}
 \author{David Zueco}
 \author{Peter H\"anggi}
 \author{Sigmund Kohler}
 \affiliation{Institut f\"ur Physik, Universit\"at Augsburg,
         Universit\"atsstra{\ss}e~1, D-86135 Augsburg, Germany}
 \date{\today} 
%
\begin{abstract}

We propose a scheme for monitoring coherent quantum dynamics with good
time-resolution and low backaction, which relies on the response of the
considered quantum system to high-frequency ac driving.  An
approximate analytical solution of the corresponding quantum master
equation reveals that the phase of an outgoing signal, which can
directly be measured in an experiment with lock-in
technique, is proportional to the expectation value of a particular
system observable. This result is corroborated by the numerical solution of
the master equation for a charge qubit realized with a Cooper-pair
box, where we focus on monitoring coherent oscillations.

\pacs{
42.50.Dv,   
03.65.Yz,   
03.67.Lx,   
85.25.Cp    
}
\end{abstract}
\maketitle


An indispensable requirement for a quantum computer is the
readout of its state after performing gate operations. For that
purpose it is sufficient to distinguish between two possible
logical states. At the same time, it is desirable to demonstrate the
coherence of time evolution explicitly. For solid-state qubits, this  
has been accomplished by Rabi-type experiments \cite{Vion2002a, 
Chiorescu2003a}. In general the qubit state measurement is
destructive, so that an interference pattern emerges only after a
number of experimental runs.  Certainly, it would be preferable to
observe signatures of coherent dynamics already in a \textit{single} run.

Both the charge and the flux degree of freedom of superconducting qubits
can be measured by coupling the qubit to a low-frequency ``tank'' 
circuit that is excited resonantly \cite{Grajcar2004a, Sillanpaa2005a}.
In doing so one makes use of the fact that the resonance frequency of the
slow oscillator depends on the qubit state which, in turn, 
influences the phase of the oscillator response 
\cite{Grajcar2004a, Sillanpaa2005a, Johansson2006a}.  The drawback
of this scheme, however, is that the coherent qubit dynamics is
considerably faster than the driving.  Thus, one can
only observe the time-average of the qubit state, but not time-resolve
its dynamics.
Measuring the qubit by driving it at resonance is possible as well
\cite{Greenberg2005a}.  This however induces Rabi oscillations, making
the qubit dynamics differ significantly from the undriven case
\cite{Liu2006a, Wilson2007a}.

Here, by contrast, we propose to probe the qubit by a weak
high-frequency driving that directly acts upon the qubit  
without a tank circuit being present.  We find that the resulting
outgoing signal possesses sidebands which are related to a phase
shift and  demonstrate that the latter is related to a qubit
observable. Validating this relation numerically for a Cooper-pair
box, we show that the underlying measurement scheme principally
enables monitoring the coherent qubit dynamics experimentally in a
single run with good time-resolution and fidelity, whereas the
backaction on the qubit, induced by the driving, stays at a  tolerable
level. 


\section{Dissipative quantum circuit}
\begin{figure}[b]
\includegraphics[width=.9\linewidth]{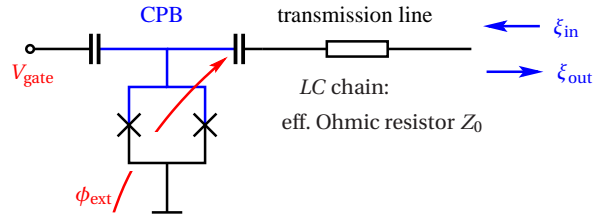}
\caption{
(color online)
Cooper pair box (CPB) coupled to a transmission line with ohmic
effective impedance $Z_0$.}\label{fig:setup}
\end{figure}

Although later on we focus on the dynamics of a superconducting charge
qubit as sketched in Fig.~\ref{fig:setup}, our measurement scheme is
rather generic and can be applied to any open quantum system.  We
employ the system-bath Hamiltonian \cite{Leggett1987a, Hanggi1990a,
Makhlin2001a}
\begin{equation}\label{eq:1}
H = H_0 
+\sum_k\Big(\frac{p_k^2}{2L_k}+\frac{(q_k -\lambda_kC_kQ)^2}{2C_k}\Big) ,
\end{equation}
where $H_0$ denotes the system Hamiltonian and $Q$ is a
system operator.  The bath is modelled by $LC$ 
circuits with charges $q_k$ and conjugate momenta $p_k$, where $C_k$
and $L_k$ are effective capacitances and inductances, respectively, and
$\lambda_k$ are the corresponding coupling constants.
%
%
It is convenient to introduce the spectral density
$I(\omega) = \frac{\pi}{2} \sum_k \lambda_k^2(C_k/L_k)^{1/2} 
\delta(\omega-\omega_k)$ which we assume to be ohmic, i.e.\
$I(\omega) = \omega Z_0$ with an effective impedance $Z_0$
\cite{Yurke1984a, Devoret1995a, Makhlin2001a}.
By standard techniques we obtain the
Bloch-Redfield master equation for the reduced system density operator 
$\rho$ in the weak-coupling limit \cite{Blum1996a},
\begin{equation}
\label{master}
\dot\rho =
-\frac{i}{\hbar}[H_0,\rho] 
- \frac{1}{\hbar}
  [ Q, [\hat Q, \rho] ]
- i\frac{Z_0}{\hbar} [ Q, [ \dot Q, \rho ]_+ ] ,
\end{equation}
with the anticommutator $[A,B]_+ = AB+BA$ and the operators
$\dot Q = i[H_0,Q]/\hbar$ and
\begin{equation}
\hat Q = \frac{1}{\pi} \int_0^\infty d\tau \int_0^\infty d\omega\,
   S(\omega) \cos(\omega\tau) \tilde Q(-\tau) .
\end{equation}
Here, $S(\omega) = I(\omega)\coth(\hbar\omega/2k_{\rm B} T)$ is the
Fourier transform of the symmetrically ordered equilibrium
correlation function $\frac{1}{2} \left\langle[\xi_\text{in}(\tau),
\xi_\text{in}(0)]_+ \right\rangle_\text{eq}$ at temperature $T$ with
regard to the collective bath coordinate $\xi_\text{in} = \sum_k
\lambda_k q_k$.  The notation $\tilde X(t)$ is a shorthand for the
Heisenberg operator $U_0^\dagger(t) X U_0(t)$, where $U_0$ is the
system propagator.

In order to relate the quantum dynamics of the central circuit to
the response via the transmission line, we employ the input-output
formalism \cite{Gardiner1985b} which is an established tool in quantum
optics and has also been used in quantum circuit theory
\cite{Johansson2006a}. It starts from the Heisenberg equation of
motion for the environmental mode $k$ which reads $\ddot q_k+\omega_k^2
q_k = \omega_k^2 Q$, where $\omega_k=(L_kC_k)^{-1/2}$ denotes the
angular frequency of mode $k$.  Owing to its linearity, this equation
of motion can be solved formally.  Inserting the obtained
solution into the Heisenberg equations of motion for the system
operators, one arrives at  the so-called quantum Langevin equation
\cite{Ford1988a, Hanggi2005a}. For an ohmic environment, the latter
possesses the inhomogeneity $\xi_\text{in}(t) - Z_0\dot Q(t)$, where
the noise operator $\xi_\text{in}$ is fully determined by the
correlation functions given above. 

Alternatively, one can write the quantum Langevin equation in terms of
the outgoing fluctuations \cite{Gardiner1985b}.  The result differs
only by the sign of the dissipative term, so that the
inhomogeneity now reads $\xi_\text{out}(t) + Z_0\dot Q(t)$.  The
difference between these two equations relates the input and the
output fluctuations via
\begin{equation}\label{in-out}
\xi_\text{out} - \xi_\text{in}
= -2  Z_0 \dot Q
= -\frac{2i  Z_0}{\hbar}\big[H_0,Q\big] ,
\end{equation}
which is a cornerstone of the input-output formalism
\cite{Gardiner1985b} and holds for any weakly coupled $\xi_\text{in}$.
We used that for weak dissipation, $\dot Q \approx
i\big[H_0,Q\big]/\hbar$ is essentially bath-independent.

\section{Response to high-frequency driving}
We next probe the system by driving it via the transmission line with
an ac signal $A\cos(\Omega t)$ that also couples to the system
operator $Q$.  Then, the Hamiltonian acquires an additional term: $H_0
\to H_0 + QA\cos(\Omega t)$, and the master equation~\eqref{master}
changes accordingly.  For the input $\xi_\text{in}$, this corresponds
to one coherently excited incoming mode such that
$\langle\xi_\mathrm{in}(t)\rangle = A\cos(\Omega t)$, while the
r.h.s.\ of the input-output relation \eqref{in-out} remains unchanged.

Because the driving must not significantly alter the system dynamics,
we assume that the amplitude $A$ is sufficiently small, so that the
driving can be treated perturbatively. This yields the ansatz
$\rho(t) = \rho_{0}(t) + \rho_{1}(t)$, 
where $\rho_{0}(t)$ is the unperturbed state.  To lowest order in $A$,
$\rho_1$ obeys $\dot\rho_1 = \mathcal{L}_0\rho_1
-\frac{i}{\hbar} A[Q,\rho_0]\cos(\Omega t)$, where $\mathcal{L}_0$
denotes the superoperator on the r.h.s.\ of Eq.~\eqref{master}.
This linear inhomogeneous equation of motion can be solved formally in
terms of a convolution between the propagator of the undriven system
and the inhomogeneity.  If the driving frequency $\Omega$ is much
larger than all relevant system frequencies, one may
separate time scales to obtain
\begin{equation}\label{eq:4}
\rho(t) = \rho_0(t)
  -i\frac{A}{\hbar\Omega}[Q,\rho_0(t)]\sin(\Omega t) \, ,
\end{equation}
which identifies $eA/\hbar \Omega$ as the necessarily small
perturbation parameter.
Together with the input-output relation~\eqref{in-out} 
this solution allows us to compute the response of the system. 

In an experiment it is possible to employ a lock-in 
technique with the incoming signal providing the reference oscillator.
This singles out the high-frequency components of the outgoing signal, 
which correspond to the second term of the density
operator~\eqref{eq:4} and read
\begin{equation}
\label{eq:5}
\langle\xi_\text{out}(t)\rangle
= A \cos(\Omega t) + \frac{2 A Z_0}{\hbar^2\Omega} 
 \left\langle [[H_0,Q],Q ] \right\rangle_0\sin(\Omega t)
 ,
\end{equation}
where $\langle\ldots\rangle_0 = \tr[\rho_0(t)\ldots]$ refers
to the undriven dynamics. 
Writing next $\langle\xi_\text{out}(t)\rangle = A'
\cos[\Omega t - \phi_\mathrm{hf}^0(t)]$, we find the central expression 
\begin{equation}\label{eq:6}
\phi_\mathrm{hf}^0(t)
= \frac{2 Z_0}{\hbar^2\Omega}
 \left\langle [[H_0,Q],Q ] \right\rangle_0 ,
\end{equation}
which relates a small phase shift  $\phi ^0_\mathrm{hf}(t)$ between
the input and the output signal to a hermitian system
observable describing the unperturbed 
low-frequency system dynamics.  This means that the
\textit{time-resolved} evolution of the open quantum system can be
monitored  in a single run by continuously
measuring the phase shift $\phi ^0_\mathrm{hf}(t)$ with appropriate
experimental techniques. In this connection, Eq.~\eqref{eq:6}
represents the basis for our proposed measurement scheme.   
Below we will explore its validity and limitations for a
specific system by comparing  the phase of the output
$\langle\xi_\text{out}(t)\rangle$ with the expectation value
$\left\langle [[H_0,Q],Q ] \right\rangle_0$, both computed from the
numerical solution of the master equation~\eqref{master} in
the presence of ac driving.

\section{Monitoring coherent qubit dynamics}
A particular case of a quantum circuit which recently attracted much
interest is a Cooper pair box (CPB) which is sketched in
Fig.~\ref{fig:setup} and described by the Hamiltonian
\begin{equation}
\label{CPB}
H_0^\mathrm{CPB}
= 4 E_\mathrm{C}(\hat N-N_g)^2 -\frac{E_\mathrm{J}}{2}
  \sum_{N=-\infty}^\infty (|N{+}1\rangle\langle N|+\text{h.c.}) ,
\end{equation}
where $N$ is the number of excess Cooper pairs in the box, so that the
charge operator reads $Q=2e\hat N  = 2e\sum_N N|N\rangle\langle N|$.
The charging energy $E_\mathrm{C}$ is determined by various
capacitances, while the scaled gate voltage $N_g$ and the effective
Josephson energy $E_\mathrm{J}$ are controllable.
If the charging energy is sufficiently large and $N_g\approx 1/2$,
only the two charge states $|0\rangle \equiv |{\downarrow}\rangle$ and
$|1\rangle\equiv |{\uparrow}\rangle$ matter and form a
qubit \cite{Makhlin2001a, Vion2002a, Grajcar2004a}
described by the Hamiltonian
\begin{equation}
H_0^\mathrm{qb}
=
-\frac{1}{2}E_\mathrm{el}\sigma_z  -\frac{1}{2}E_\mathrm{J}\sigma_x,
\end{equation}
where the Pauli matrices $\sigma_i$
are defined in the qubit subspace and $E_\mathrm{el} =
4E_\mathrm{C}(1-2N_g)$. The qubit energy splitting reads
$\hbar\omega_\mathrm{qb} = (E^2_\mathrm{el} + E^2_\mathrm{J})^{1/2}$.
Moreover, $Q_\mathrm{qb} = e\sigma_z$
while by virtue of relation~\eqref{eq:6} the phase of the output
signal is linked to the qubit observable $\sigma_x$ according to
\begin{equation}\label{eq:7}
\phi_\mathrm{hf}^0(t)
= - \frac{ 4e^2 Z_0 E_\mathrm{J}}{\hbar^2 \Omega}
 \left\langle \sigma_x (t) \right\rangle_0 \, .
\end{equation}
This means that the high-frequency component of $\langle\dot
Q\rangle$, which is manifest in the phase of the outgoing signal
\eqref{in-out}, contains information about the low-frequency qubit
dynamics in terms of the unperturbed $\langle\sigma_x\rangle_0$.

\begin{figure}[tb]
\includegraphics{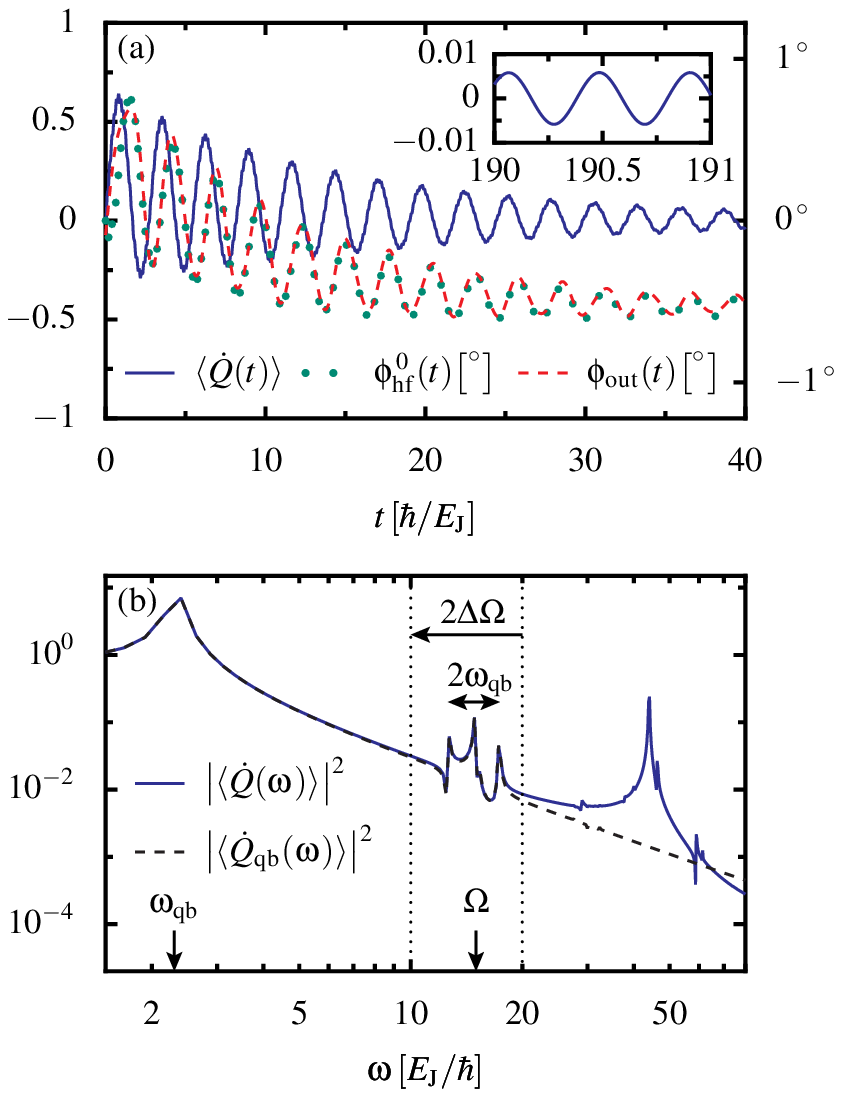}
\caption{(color online) 
Decaying qubit oscillations with initial state $|{\uparrow}\rangle$ in
a weakly probed CPB with 6 states 
for $\alpha =Z_0 e^2/\hbar =0.08$, $A=0.1 E_\mathrm{J}/e$,
$E_\mathrm{C}=5.25 E_\mathrm{J}$ and $N_g=0.45$, so that
$E_\mathrm{el} = 2.1 E_\mathrm{J}$ and $\omega_\mathrm{qb} = 2.3
E_\mathrm{J}/\hbar$.
(a) Time evolution of the measured difference signal $
\langle\dot Q \rangle \propto \langle\xi_\mathrm{out}\rangle -
\langle\xi_\mathrm{in}\rangle$
(in units of $2 e E_\mathrm{J} /\hbar$)
of the full CPB and its lock-in amplified phase $\phi_\text{out}$
(frequency window $\Delta\Omega=5 E_\mathrm{J}/\hbar$), compared to
the estimated phase $\phi _\mathrm{hf}^0
\propto \langle\sigma_x\rangle_0$ in the qubit approximation.
The inset resolves the underlying small rapid oscillations with
frequency $\Omega=15 E_\mathrm{J}/\hbar$ in the long-time limit.  
(b) Power spectrum of $ \langle\dot Q\rangle$ for the
full CPB Hamiltonian (solid) and for the two-level approximation (dashed).
}\label{fig:oscillation}
\end{figure}%
%
We now turn to the question how relation \eqref{eq:7} allows one to
retrieve information about the coherent qubit dynamics in an
experiment.
Figure~\ref{fig:oscillation}(a) shows the time evolution of the
expectation value $\langle\dot Q (t)\rangle$ for the initial state  
$|{\uparrow}\rangle \equiv |1\rangle$, obtained via numerical
integration of the master equation~\eqref{master} with the full
Cooper-pair box Hamiltonian~\eqref{CPB} in the presence of the ac
driving which in principle may excite higher states.
The driving, due to its rather small amplitude, is barely noticeable on
the scale chosen for the main figure, but only on a refined scale for
long times; see inset of Fig.~\ref{fig:oscillation}(a).
This already insinuates that the backaction on the dynamics is weak. 
In the corresponding power spectrum of $\langle\dot Q \rangle$
depicted in Fig.~\ref{fig:oscillation}(b), the driving is nevertheless
reflected in sideband peaks at the frequencies $\Omega$ and $\Omega 
\pm\omega_\mathrm{qb}$. In the time domain these peaks correspond to a
signal $\cos[\Omega t -\phi_\mathrm{out} (t)]$.  Moreover,
non-qubit CPB states leads to additional peaks
at higher frequencies, while their influence at frequencies $\omega
\lesssim \Omega$ is minor.
Experimentally, the phase $\phi_\mathrm{out}(t)$ can be retrieved by
lock-in amplification of the output signal, which we mimic numerically
in the following way \cite{Scofield1994a}: We only consider the
spectrum of $\xi_\text{out}$ in a window $\Omega\pm\Delta\Omega$
around the driving frequency and shift it by $-\Omega$.  The inverse
Fourier transformation to the time domain provides
$\phi_\mathrm{out}(t)$ which is expected to agree with $\phi_\text{hf}^0(t)$
and, according to Eq.~\eqref{eq:7}, to reflect the unperturbed time
evolution of $\langle\sigma_x\rangle_0$ with respect to the qubit.
Although the condition of high-frequency probing, $\Omega \gg
\omega_\mathrm{qb}$, is not strictly fulfilled and despite the
presence of higher charge states, the lock-in amplified
phase $\phi_\text{out} (t)$ and the predicted phase $\phi ^0_\text{hf}
(t)$ are barely distinguishable for an appropriate choice of
parameters as is shown in Fig.~\ref{fig:oscillation}(a).

\begin{figure}[tb]
\includegraphics{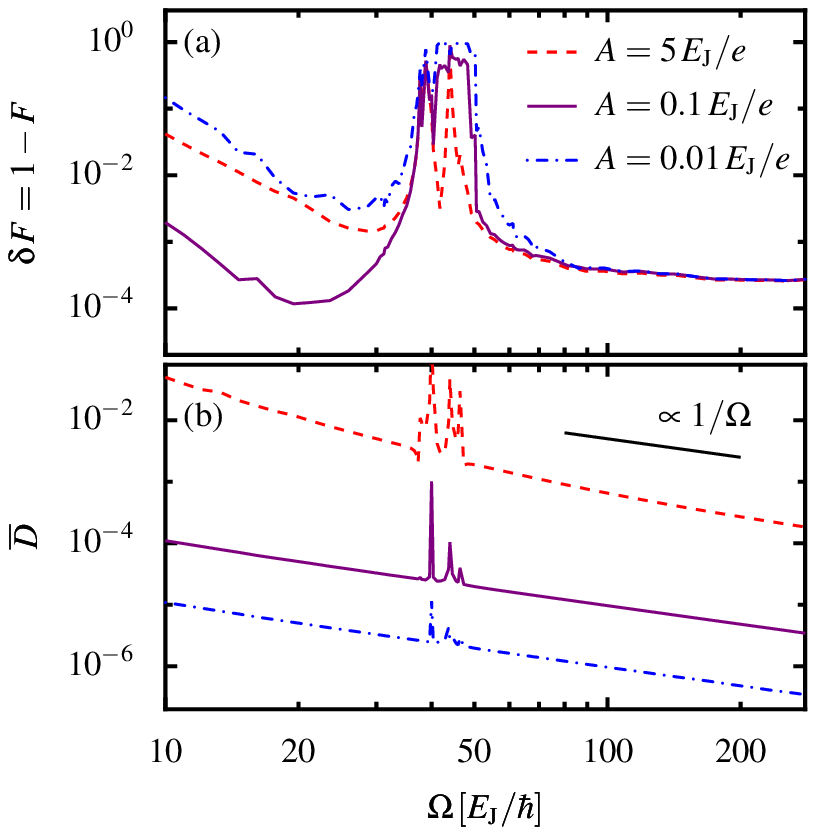}
\caption{(color online) (a) Fidelity defect $\delta F=1-F$ and (b)
time-averaged trace distance between the driven and the undriven
density operator of the CPB for various driving amplitudes as a
function of the driving frequency.
All other parameters are the same as in Fig.~\ref{fig:oscillation}.
}
\label{fig:fidelity} 
\end{figure}

In order to quantify this agreement, we introduce the measurement
fidelity $F = ( \phi_\mathrm{out} , \langle\sigma_x\rangle_0 )$, where
$(f,g) = \int dt\, fg / (\int dt\, f^2 \int dt\, g^2)^{1/2}$ with time
integration over the decay duration.
Thus, the ideal value $F=1$ is assumed if $\phi_\mathrm{out}(t)$ and
$\langle\sigma_x(t)\rangle_0$ are proportional to each other, i.e.\
if the agreement between the measured phase and the unperturbed
expectation value $\langle \sigma_x \rangle _0$ is perfect.
Figure~\ref{fig:fidelity}(a) depicts the fidelity as a function of the
driving frequency.  As expected, whenever non-qubit CPB states are excited
resonantly, we find $F\ll 1$, indicating a significant population of
these states.
Far-off such resonances, the fidelity increases with the
driving frequency $\Omega$.  A proper frequency lies in the middle
between the qubit doublet and the next higher state.  In the present
case, $\Omega \approx 15 E_\text{J}/\hbar$ appears as a good choice.
Concerning the driving amplitude, one has to find a compromise,
because as $A$ increases, so does the phase contrast of the outgoing
signal \eqref{eq:5}, while the driving perturbs more and more the
low-frequency dynamics.  For the frequency chosen above, the fidelity
is best in the range
$A = 0.1$--$1 \hbar\omega_\mathrm{qb}/e$.  This corresponds to
$e A/\hbar\Omega \approx 10^{-2}$--$10^{-3}$, which
justifies our perturbative treatment.

\section{Measurement quality and backaction}
For any quantum measurement, one has to worry about backaction
on the system in terms of decoherence.  In our
measurement scheme, decoherence plays a particular role, because both
the driving and the ohmic environment couple to the CPB via the same
mechanism.  This is reflected by the fact that the predicted phase
\eqref{eq:7} is proportional to the dimensionless dissipation strength
$\alpha = Z_0 e^2/\hbar$.  However, $\alpha$ should not exceed
$0.1$ in order to preserve a predominantly coherent time evolution,
which also means that our measurement is weak, but destructive.
The condition $\alpha\lesssim 0.1$ together with the above conditions
on the driving amplitude and frequency provides phase shifts
$\phi_\mathrm{out}$ of the order $1^{\circ}$, which is small but still
measurable with present technologies.
The additional decoherence due to the driving, by contrast, is not
noticeable.
This is in agreement with the first-order result $\tr(\rho^2) =
\tr(\rho_0^2)$ which follows from Eq.~\eqref{eq:4}.

In order to investigate to what extent the driving affects the quantum
state of the CPB, we compute the trace distance $D(t) = \frac{1}{2}
\tr|\rho(t)-\rho_0(t)|$ \cite{Nielsen2000a} between the density
operators of the driven system $\rho(t)$ and the undriven reference
$\rho_0(t)$. Its time average $\overline D$ quantifies the perturbation
due to the driving.
Figure~\ref{fig:fidelity}(b) indicates that $\overline D\propto A/\Omega$,
unless the driving is in resonance with higher levels.  This confirms
the picture drawn by studying the measurement fidelity $F$.
For practically all parameters used in Fig.~\ref{fig:fidelity}, we
found that the total population of levels outside the qubit doublet is
always less than 0.1\%.  The only exception occurs again in the case
of resonances with non-qubit CPB levels.  Far from these resonances,
the system is faithfully described with only the qubit levels.

In our investigations, we have not considered excitations of
quasiparticles which are relevant once the driving frequency becomes
of the order of the gap frequency of the superconducting material.
Thus, for an aluminum CPB our model is valid only for
$\Omega \lesssim 100\,\mathrm{GHz}$.  Since a typical
Josephson energy is of the order of some GHz, a driving frequency
$\Omega \approx 20 E_\mathrm{J}/\hbar$ is still within this range
while it provides already a good measurement quality.

\section{Conclusions}
We have proposed a method for the time-resolved monitoring of the
dynamics of a quantum system coupled to a dissipative environment.
The crucial requirement for this is the possibility to drive the
system coherently via one environmental high-frequency mode
accompanied by measuring the phase of the outgoing signal via lock-in
techniques.  By analyzing the  high-frequency response, we have found
that the phase of the output signal is related to a particular system
observable.
We have substantiated this relation by computing both quantities
numerically for a charge qubit implemented with a Cooper-pair box.
For decaying coherent oscillations, we have demonstrated experimental
feasibility on condition that the coupling to the environment is not
too weak and that the driving frequency exceeds the qubit splitting
and is off resonance with higher levels.  Then the measurement
fidelity is rather good, while the low-frequency qubit dynamics is
almost not affected by the driving and transitions to higher levels do
not play a relevant role.
The implementation of our scheme will enable the demonstration of
quantum coherence of solid-state qubits in single-shot experiments.

This work has been supported by the DFG through SFB 631 and by the
German Excellence Initiative via ``Nanosystems Initiative Munich
(NIM)''.


%

\end{document}